\documentclass[a4paper,twocolumn,superscriptaddress,prl, aps, floatfix ]{revtex4-1}

\usepackage{epsfig}
\usepackage{verbatim}	 
\usepackage{amssymb}
\usepackage{bm}
\usepackage{amsmath}
\usepackage{psfrag}
\usepackage{color}
\usepackage[latin1]{inputenc}
\usepackage[english]{babel}
\usepackage{graphicx}
\usepackage{braket}
\usepackage{eucal}
\usepackage{natbib}
\usepackage{hyperref}
\usepackage{lipsum}
\usepackage{mathtools}
\usepackage{multirow}
\usepackage{cleveref}
\usepackage[colorinlistoftodos]{todonotes}

\addto\captionsenglish{}

\begin{document}
\title{Modeling echo chambers and polarization dynamics in social networks }

\author{Fabian Baumann}
\thanks{Corresponding author: fabian.olit@gmail.com}
\affiliation{Institute for Physics, Humboldt-University of Berlin, Newtonstra\ss e 15, 12489 Berlin, Germany}

\author{Philipp Lorenz-Spreen}
\affiliation{Center for Adaptive Rationality, Max Planck Institute for Human Development, Lentzeallee 94, 14195 Berlin, Germany}

\author{Igor M. Sokolov}
\affiliation{Institute for Physics, Humboldt-University of Berlin, Newtonstra\ss e 15, 12489 Berlin, Germany}
\affiliation{IRIS Adlershof, Humboldt-University of Berlin, Newtonstra\ss e 15, 12489 Berlin, Germany}

\author{Michele {Starnini}}
\thanks{Corresponding author: michele.starnini@gmail.com}
\affiliation{ISI  Foundation,  via  Chisola  5,  10126  Torino, Italy}

\begin{abstract}
Echo chambers and opinion polarization 
recently quantified in several sociopolitical contexts and across different social media,
raise concerns on their potential impact on the spread of misinformation and on openness of debates. 
Despite increasing efforts, the dynamics leading to the emergence of these phenomena stay unclear. 
We propose a model that introduces the dynamics of radicalization, as a reinforcing mechanism 
driving the evolution to extreme opinions from moderate initial conditions. 
Inspired by empirical findings on social interaction dynamics, we consider agents characterized by 
heterogeneous activities and homophily.
We show that the transition between a global consensus and emerging 
radicalized states is mostly governed by  social influence and by the controversialness of the topic discussed. 
Compared with empirical data of polarized debates on Twitter, the model qualitatively reproduces the observed relation between users' engagement and opinions, 
as well as opinion segregation in the interaction network.
Our findings shed light on the mechanisms that may lie at the core of the emergence of echo chambers and polarization in social media.
\end{abstract}


\maketitle



The participatory character of political debates on online social media \cite{gil2011mediating} adds degrees of freedom to the
self-organisation of public opinion formation \cite{moussaid2013social}.
The low cost for engagement and the distributed architecture of communication infrastructures have increased interaction rates \cite{lorenz2019accelerating}, and lowered barriers due to geographical distance or social status \cite{internetworldstats}. 
Within the traditional models based on constructive opinion dynamics 
\cite{degroot1974reaching}, such unrestricted 
modes of interaction would eventually lead to a consensus, even on
controversial issues. 

However, this prediction is not always confirmed empirically: 
 heterogeneous and bimodal distributions of opinions have been measured in political surveys \cite{glaeser2006myths, baldassarri2008partisans}, especially on {controversial} issues, like abortion or global warming \cite{dimaggio1996have, mccright2011politicization}.
We refer to situations in which the opinion distribution is characterized by two well-separated peaks around the neutral consensus as \emph{polarization}.
In social media, such polarized communication networks were observed in controversial debates ranging from political orientation~\cite{adamic2005political, CON11}, US and French presidential 
elections~\cite{Hanna:2013:PAP:2508436.2508438}, or street protests~\cite{Borge-Holthoefer15}. 
If segregation in the opinion space is reflected in interactions among users, \emph{echo chambers} emerge: situations in which one's opinion resonates with those of ones' social contacts~\cite{garrett2009echo}. 
Echo chambers have been quantified in several controversial debates on different social media platforms~\cite{Garimella:2018:PDS:3178876.3186139,DBLP:journals/corr/VicarioVBZSCQ16,Cota2019} and may be related to the spread of misinformation~\cite{del2016spreading}. 

The contradiction between empirical observations and predictions of classical models of opinion dynamics question the mechanisms that drive opinion polarization and the formation of echo chambers. 
Previous modeling approaches describe segregation of opinions by influence based on similarity in the opinion space, either through a confidence bound \cite{HEG02, deffuant2000mixing} or through homophily \cite{axelrod1997dissemination}, the preference of agents to interact with similar individuals \cite{mcpherson2001birds, bessi2016homophily}.
Another class of models describe polarization by introducing repulsive interactions, in which users reject opinions that differ from their own \cite{macy2003polarization, martins2010mass, flache2011small}, but this mechanism combined with homophily leads to a decrease of polarization \cite{maes2015will, flache2017models}.
Although, echo chambers and modular network structures were previously also related to 
homophily \cite{starnini2016emergence, sasahara2019inevitability}, several empirical 
features of social networks characterized by echo chambers 
\cite{Garimella:2018:PDS:3178876.3186139, DBLP:journals/corr/VicarioVBZSCQ16,  
Cota2019, sasahara2019inevitability} and their relations 
to opinion polarization have not been addressed within a unified modeling framework.


In this Letter, we propose a simple model of opinion dynamics that can capture this 
relation and reproduce two empirical features frequently observed in polarized social networks:
i) more active users, stronger engaged in social interactions, tend to show more extreme opinions,
and ii) the opinion expressed by a user and those expressed by his/her neighbors in the social interaction network are similar. 
The model introduces a mechanism by which agents sharing similar 
opinions can mutually reinforce each other 
and move towards more extreme views, 
thus describing a \emph{radicalization dynamics}, also known as group polarization in social psychology \cite{isenberg1986group, myers1976group}. 
Alongside, opinion states are coupled to the underlying time-varying network of social interactions 
by homophily \cite{PhysRevE.74.056108,PhysRevE.78.016103}.
While the convergence toward a global consensus is retained, the introduction of opinion reinforcement and homophily leads to the emergence of metastable polarized states.
The transition between consensus and radicalization dynamics is analytically characterized on the basis of the interplay between social influence and the controversialness  of the topic discussed.

Let us consider a system of $N$ agents. Each agent $i$ is characterized by a one-dimensional opinion variable $x_i(t)\in [-\infty, +\infty]$. 
The sign of $x_i$, $\sigma(x_i)$, describes the agent's 
qualitative stance towards a binary issue of choice (e.g. the preference 
between two candidates). 
The absolute value of $x_i$, $|x_i|$, quantifies the strength of this opinion,  or the conviction,  with respect to one of the sides: the larger $|x_i|$, the more extreme the stance of agent $i$.
We assume that the opinion dynamics is solely driven by the interactions among agents, and describe it by a system of $N$ coupled ordinary differential equations,
\begin{equation}\label{eq:model}
  \dot{x}_i = -x_i + K \sum_{j=1}^N A_{ij}(t) \tanh(\alpha x_j),
\end{equation}
where $K > 0$ denotes the social interaction strength among agents and $\alpha>0$ controls the shape of the sigmoidal influence function taken to be $\tanh(\alpha x)$. 
Its odd non-linear shape guarantees that 
an agent $j$ influences others in the direction of its own opinion's sign, $\sigma(x_j)$, that this influence increases monotonically with the agent's conviction $|x_j|$, and that the social influence of extreme opinions is capped, as suggested by experimental findings \cite{jayles2017social}. 
Similar odd and tunable functions have previously been used to model non-linear gain functions in mean-field models of neural systems, to study chaotic dynamics \cite{sompolinsky1988chaos} or the effects of gain on attention and learning \cite{eldar2013effects}.

According to Eq.~\eqref{eq:model}, the opinion of an agent $i$ changes depending on the aggregated inputs from his/her neighbors.
This mechanism builds on the idea of informational influence theory \cite{10.1371/journal.pone.0074516} for the phenomenon of group polarization, where agents with moderate opinions may become extreme while interacting in a group \cite{isenberg1986group}.
The neighbors of agent $i$ are determined by the temporal adjacency matrix $A_{ij}(t)$, with $A_{ij}(t)=1$ if there is input from agent $j$ to agent $i$ at time $t$, $A_{ij}(t)=0$ otherwise.
Information flow on social media is in general asymmetric with the degree of asymmetry depending on the social media platform under consideration. 
While social interactions are initiated asymmetrically, they may easily generate feedback.
Hence, we consider directed interactions, that may be reciprocated with a certain probability $r$.
When agent $i$ establishes a connection to another agent $j$, agent $j$ will update its opinion, but agent $i$ will do the same only if the interaction is reciprocated.

For a reciprocal interaction between $i$ and $j$, 
we distinguish two fundamentally different situations, depending on the signs of their opinions $\sigma(x)$.
If the agents share the same stance ($\sigma(x_i)$ = $\sigma(x_j)$),
the interaction will cause an increase of both convictions and hence reinforce opinions, 
a mechanism we refer to as \emph{radicalization} dynamics.
On the contrary, for opposing stances ($\sigma(x_i)$ = $-\sigma(x_j)$), opinions tend to converge. 
Note that we model opinion dynamics as a purely collective, 
self-organized process without any intrinsic individual preferences. 
Hence, the opinions of agents lacking social interactions
decay towards the neutral state.

The parameter $\alpha$ tunes the degree of non-linearity between an agent's opinion and the social influence s/he exerts on others.
For small $\alpha$, the social influence of moderate  individuals   
on other peers is weak. 
For large $\alpha$, by contrast, even agents with moderate
opinions can already exert a strong social influence on others. 
The limit of $\alpha \to \infty$, with $\tanh(\alpha x) \rightarrow \sigma(x)$,
corresponds to a binary vote of maximal social influence.
Therefore, the parameter $\alpha$ is interpreted as the \emph{controversialness} of the issue.
Empirically, it has been shown that controversy is an important factor driving the 
emergence of polarization and echo chambers in debates on online social 
media \cite{Garimella:2018:QCS:3178568.3140565}.


The contact pattern, sustaining the opinion dynamics, represents social interactions which are known to evolve in time \cite{Barabasi:2005uq} and is coded in $A_{ij}(t)$.
Following empirical observations, we model the interaction dynamics as an
activity-driven (AD) temporal network
\cite{perra2012activity,PhysRevE.87.062807,Moinet2015,PhysRevLett.112.118702},
differently from previous modelling efforts proposing similar mechanisms 
on static graphs \cite{shin2010tipping}.
Here, each agent $i$ is characterized by an activity $a_i\in[\varepsilon, 1]$, representing his/her propensity to contact $m$ distinct random other agents.
Activities are extracted from a distribution $F(a)$ typically assumed to follow a power-law,  $F(a)\sim a^{-\gamma}$, 
as measured in empirical data \cite{perra2012activity, Moinet2015}. 
The set of parameters $(\varepsilon, \gamma, m)$ fully encodes the basic AD dynamics.
While in the original AD formulation agents establish connections by random uniform selection, we assume here that interactions are ruled by homophily \cite{mcpherson2001birds, bessi2016homophily}.
To this end, the probability $p_{ij}$ that an active agent $i$ will contact a peer $j$ is modeled as a decreasing function of the distance between their opinions,
\begin{equation}\label{eq:homophily}
 p_{ij} = \frac{|x_i-x_j| ^{-\beta}}{\sum_j|x_i-x_j| ^{-\beta}},
\end{equation}
where the exponent $\beta$  controls the power law decay of the connection probability with opinion distance. 
Note that the parameter $\beta$ may include various homophilic effects in interactions, both endogeneous (due to the intrinsic behavior of agents) and exogeneous (i.e. due to the algorithms of social media platforms \cite{maes2015will}). 

Here, we focus on a regime in which social interactions evolve much faster than opinions, like it is reasonable to assume for online social media.  
Attitude change, indeed, is known to be slow, especially regarding important or controversial topics \cite{krosnick1988attitude}. 
This yields a clear time-scale 
separation between the network's and opinion dynamics.

Specifically,  we choose to numerically integrate Eq. \eqref{eq:model} with $dt = 0.01$, 
while the temporal network $A_{ij}(t)$ is updated at each integration step. In the Supplementary Material~\cite{SM} we give a detailed description of our numerical algorithm. 
In the following we discuss the behavior of the model as a function of the social 
interaction strength $K$, the controversialness $\alpha$ and the homophily exponent $\beta$. 
In our simulations we use a system of $N=1000$ agents. 
We initialize the opinions uniformly spaced on the interval $x_i\in[-1, 1]$, set the AD parameters to $m=10$, $\epsilon=10^{-2}$ and 
$\gamma=2.1$ and fix the reciprocity parameter to $r=0.5$, where not differently indicated. 
In \cite{SM} we show that the obtained results are robust with respect to $r$, and asymmetric initial conditions.

\begin{figure}[tbp]
\centering
\includegraphics[width=\linewidth]{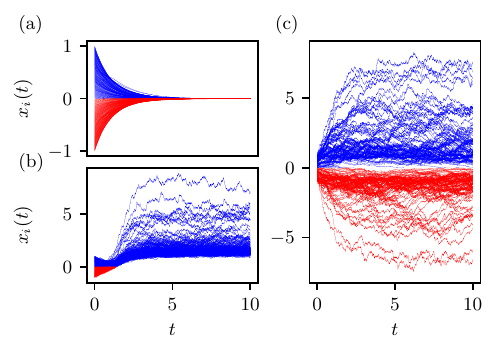}
\caption{\textbf{Temporal evolution of the agents' opinions.}
(a) Neutral consensus for which all opinions 
converge to zero ($\alpha=0.05$, $\beta=2$).
(b) (One-sided) radicalization ($\alpha=3$, $\beta=0$).
(c) Opinion polarization, in which opinions split into two opposite sides ($\alpha=3$, $\beta=3$). Social interaction strength and reciprocity were set to $K=3$ and $r=0.5$, respectively.
Positive (negative) opinions $\sigma(x_i)>0$ $(\sigma(x_i)<0)$ are colored
in blue (red). 
Note different scales on the y-axis.}
\label{fig:fig1}
\end{figure}

We identify three qualitatively different dynamical regimes. 
For small values of $K$ and $\alpha$,  a \emph{neutral consensus} is reached, in which the opinions of all agents converge towards zero, Fig.~\ref{fig:fig1}(a).
Larger values of $\alpha$ and/or $K$ destabilize this consensus and  give  rise to radicalization.
These are situations in which agents' opinions do not converge, are widely
spread, and may reach values far outside of the 
initial opinion interval. 
For such cases,  the dynamics of the system strongly depends on how active agents  choose their interaction partners. 
In the absence of homophily ($\beta=0$), when agents pick their interaction partners uniformly at random, all opinions will be directly absorbed by one of the two sides, as shown in Fig.~\ref{fig:fig1}(b). 
The introduction of homophily ($\beta>0$)  drastically changes this picture: driven by repeated interactions 
with likely-minded individuals, agents reinforce their opinions and segregate into two groups on opposite sides of the neutral consensus, as shown in Fig.~\ref{fig:fig1}(c).
In this scenario, a \emph{polarized state} characterized by a bimodal distribution of opinions emerges (see Fig. S1(b) in Ref.~\cite{SM}), 
as observed empirically \cite{BRA17a, CON11, CON12a, Cota2019, Hanna:2013:PAP:2508436.2508438, Weber2013} and in modelling studies \cite{kurahashi2016robust,banisch2019opinion}. 
The polarized state in our model is metastable and (for moderate values of $\beta$) eventually turns to a one-sided radicalized state.
Its lifetime, however, increases at least exponentially with the strength of homophily $\beta$, up to a point where the destabilization becomes numerically inaccessible (see Fig.~S2 of Ref.~\cite{SM}).

\begin{figure}[tbp]
\centering
\includegraphics[width=\linewidth]{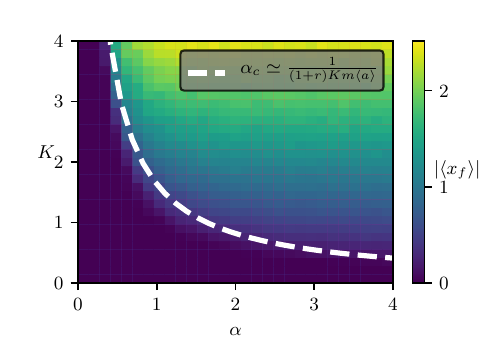}
\caption{\textbf{Transition from consensus to radicalization dynamics.} 
Absolute values of the average final 
opinions $|\braket{x_f}|$ in 
$K$-$\alpha$ phase space for $\beta=0.5$
and $r=0.5$.
In the dark region, the system approaches a neutral consensus, 
while in the colored areas the population undergoes radicalization 
dynamics which become more pronounced 
for increasing values of $K$ and/or $\alpha$ (color code).}
\label{fig:fig2}
\end{figure}

The transition from neutral consensus to radicalization is depicted in Fig.~\ref{fig:fig2} on a $K$-$\alpha$ plane, where the color encodes the 
absolute value of the final average opinion, 
$|\langle x_f\rangle| \equiv |1/N\sum_i x_i(t_{\text{final}})|$.
In the long term regime, the value of $|\langle x_f\rangle| $ identifies the transition between regions exhibiting 
a stable neutral consensus, $|\langle x_f\rangle| =0$ (dark purple), 
characterized by small values of $K$ and $\alpha$,
and regions where radicalization emerges and becomes stronger, $|\langle x_f\rangle| >0$ (color coded: blue to yellow), 
obtained for increasing $K$ and/or $\alpha$.
In the limit of fast switching interactions, this transition can be captured within a mean-field approximation. 
Neglecting homophily ($\beta = 0$) leads to the following analytical expression for the critical value of 
controversialness 
(see \cite{SM} for details),
\begin{equation}\label{eq:alpha_crit}
\alpha_c  \simeq \quad\frac{1}{(1+r)Km\braket{a}},
\end{equation}
as a function of $K$, $r$ and 
the average activity $\langle a \rangle$.
Depicted as dashed line
in Fig.~\ref{fig:fig2}, 
it still captures 
the transition 
for moderate values of $\beta$. 

 We now contrast the behavior of our model with  
three different data sets of polarized debates on Twitter, analyzed in Ref. \cite{Garimella:2018:PDS:3178876.3186139}, containing tweets on specific topics of discussion, known to be 
politically controversial: guncontrol, Obamacare, and abortion. 
The data sets have been built along two main features: i) the political orientation 
of users and ii) their social interaction network. 
Each user is characterized by his/her political leaning, 
based on established political leaning scores of various 
news organizations (e.g., nytimes.com, foxnews.com), 
ranging from very conservative to very liberal \cite{Bakshy1130}. 
Specifically, the political leaning score $x_i \in [-1,+1]$ of user $i$ 
(equivalent to $x_i$ in the model) is obtained 
by considering the set of tweets posted by user $i$ that contain links to news organizations of known political leaning.
Moreover, for each data set, the social network of interactions among the 
users is reconstructed, so that there exists a direct link from node $i$ 
to node $j$ if user $i$ follows user $j$. 
As the data sets
confirm the presence of political polarization and echo chambers in online social media 
we compare them with our model
using $r=0.65$, which 
is close to the empirically measured reciprocity values, see ~\cite{SM}
for details on the data.

The data on $x_i$ yield the distributions of expressed opinions, $P(x)$,
which show a bimodal shape across all three 
considered data sets (see Fig.~S1(a) in Ref.~\cite{SM}).
Even though the method used to infer users' opinions can differ 
(e.g. likes to Facebook pages \cite{del2016spreading}, Twitter 
hashtags \cite{Cota2019}, upvotes to Youtube videos \cite{BES16} or political leaning of media linked in tweet messages \cite{Garimella:2018:QCS:3178568.3140565}),  
the shape of the opinion distributions across diverse topics and different online social media platforms looks similar.
For sufficiently large values of $K$, $\alpha$ and $\beta$
their shapes are qualitatively well reproduced by our model, cf. Ref.~\cite{SM}.


\begin{figure}[tbp]
\centering
\includegraphics[width=\linewidth]{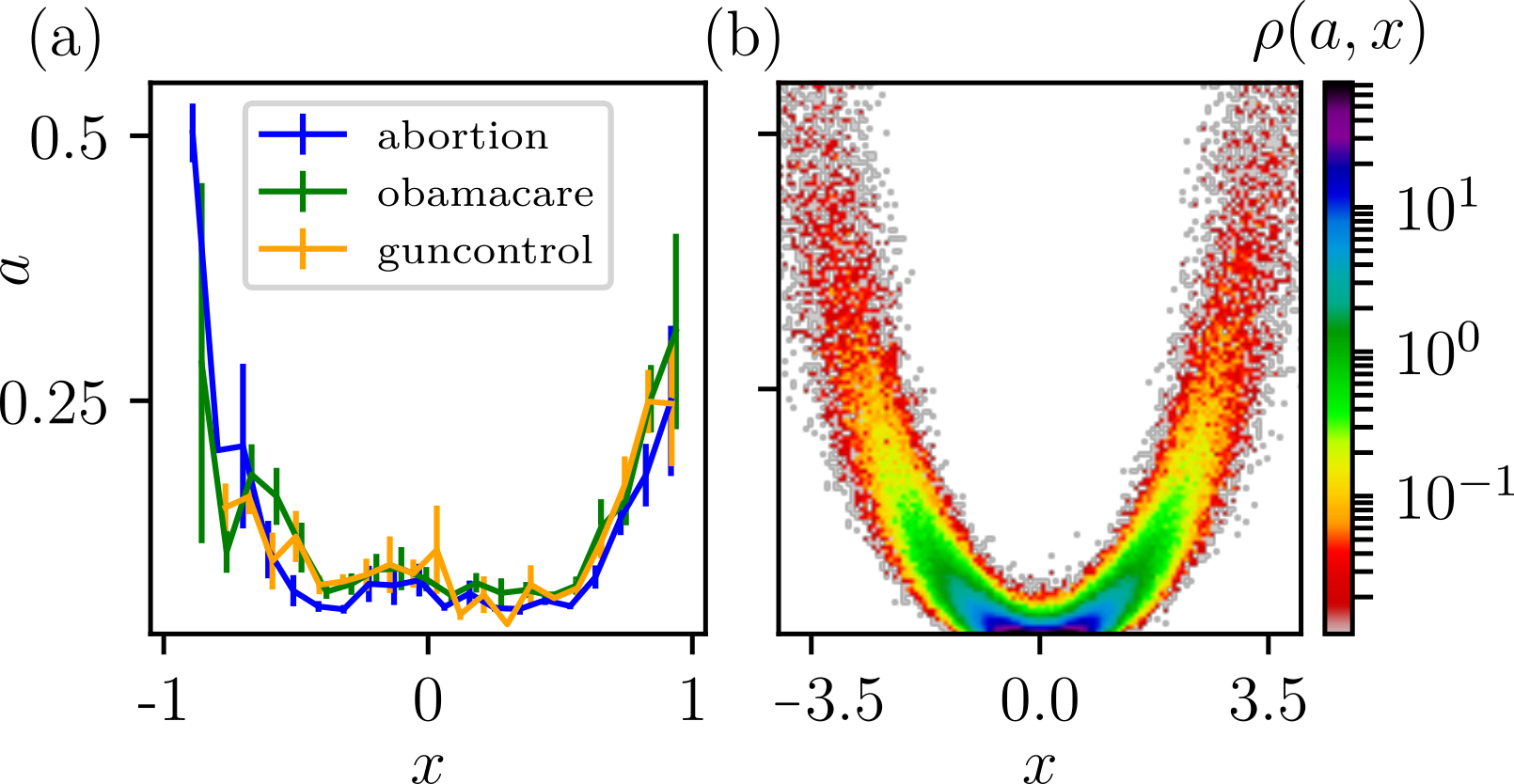}
\caption{\textbf{Activity vs. opinion.} (a) {Average} 
activity $\langle a\rangle$ 
of users as a function of their political leaning $x$, for three 
empirical data sets. (b) Activity-opinion density plot of $10^3$ 
polarized opinion states for $K=2$, $\alpha=3$, 
$\beta=1$ and $r=0.65$. The colors encode the value of $\rho(a,x)$ which is normalized with respect to $N$.}  
\label{fig:fig3}
\end{figure}

\begin{figure}[tbp]
\centering
\includegraphics[width=\linewidth]{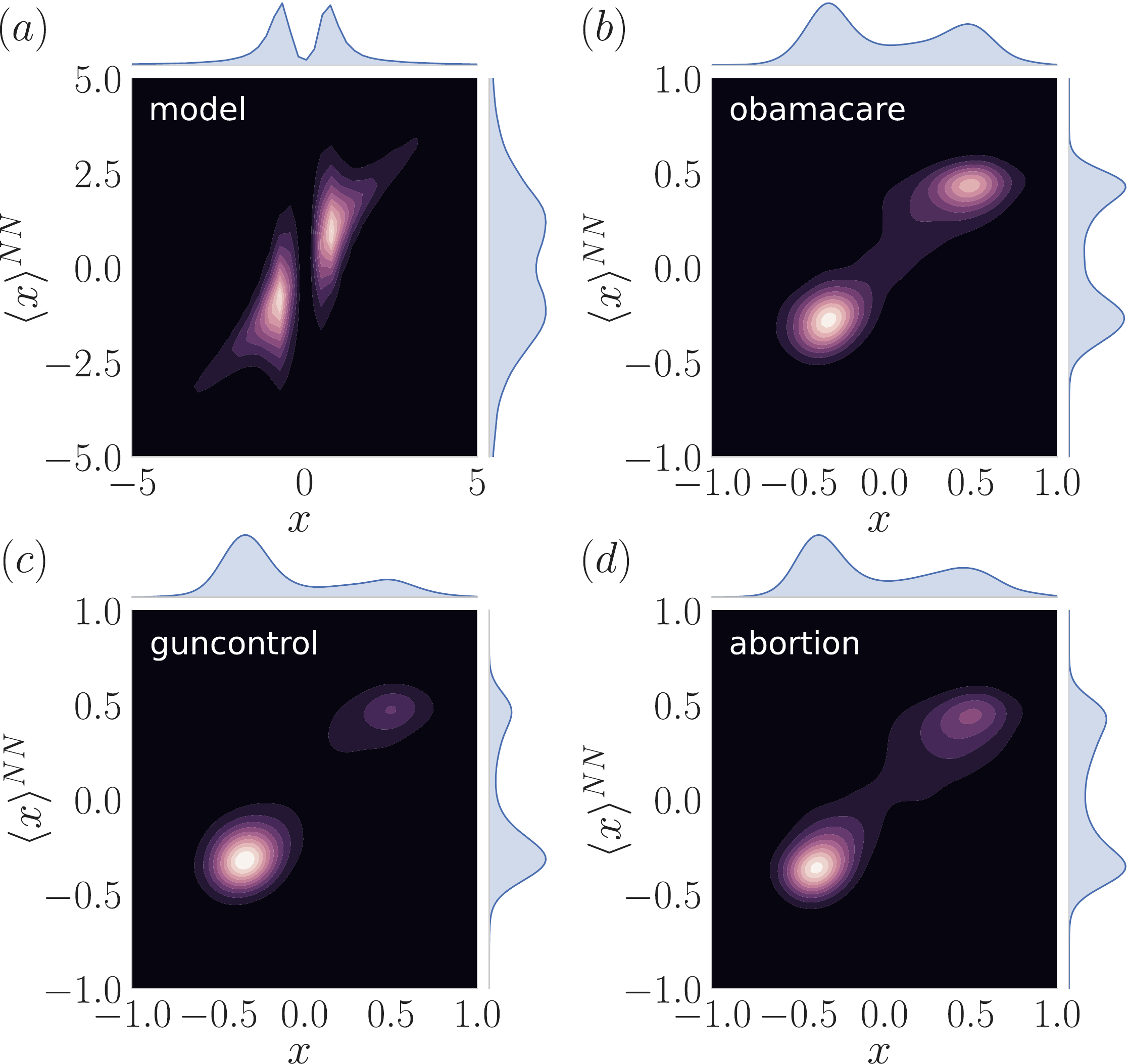}
\caption{\textbf{Echo chambers.} 
Contour maps for the average opinions of the nearest-neighbor $\langle x\rangle^{NN}$ 
against a user's opinion $x$, for 200 simulations of the radicalization model  with $K=2.5$, $\alpha=4.5$, $\beta=2$ and $r=0.65$ (a) and 
three different data sets (b-d). Colors represent the density of users: 
the lighter the larger the number of users. 
The marginal distribution of opinions, $P(x)$, and average opinions of 
the nearest-neighbor $P_{NN}(x)$ are plotted on the x- and y-axis, respectively.
}
\label{fig:fig4}
\end{figure}

A striking feature evident in different empirical data sets of polarized 
debates is a clear association between the engagement of users in the discussion 
and their convictions:
more active users 
tend to show more extreme opinions. 
For the Twitter data analyzed here,
we asses the activity of a user
as the fraction of tweets containing links to news organizations of known political leaning, a rationale derived from the original activity potential definition \cite{perra2012activity}.

Fig.~\ref{fig:fig3}(a) shows the average engagement, or activity $a$, of users as a function of their opinions $x$. 
For all three topics under consideration, the engagement rises towards the extremes of opinion space. 
It is important to note that differently defined user activity and opinion, such as the number of 
Likes to Facebook Pages tagged in different classes, shares of political content on Facebook  \cite{Bakshy1130}, 
or tweet rates of users classified according to the hashtags they use \cite{Cota2019}, give 
rise to the same functional relationship.
This characteristic U-shaped relation is well reproduced by 
our model, see Fig.~\ref{fig:fig3}(b) (shown for different parameters in Fig.~S3 in Ref.~\cite{SM}).
Within our model the finding 
suggests,
that while most users have low activities and opinions close to the neutral consensus, 
some very active users take on more extreme opinions, since their opinions are  reinforced by interactions with sufficiently likely-minded peers. 

Echo chambers are identified by the correspondence 
between the distribution of opinions in the 
population and the topology of the interaction network. 
Hence, users are more 
likely to connect to peers sharing similar opinions, which fosters information exchange among likely-minded individuals. 
On a network level,
this translates into a correlation between the opinion of a user $i$, $x_i$, 
and the average opinions of her nearest neighbors, $\langle x_i\rangle^{NN} 
\equiv {1/k_i} \sum_j a_{ij} x_j$ \cite{Cota2019}, where $a_{ij}$ represents the (static) 
adjacency matrix of the aggregated interaction network and $k_i \equiv \sum_j a_{ij}$ defines the degree of node $i$.
Fig.~\ref{fig:fig4} shows  colored contour maps of the density of users in the  $(x, \langle x\rangle^{NN})$ plane, for both empirical data and the model. 
The interaction network in Fig.~\ref{fig:fig4}(a) is obtained by aggregating 
45 snapshots of the temporal network, where the system is in a polarized state.
Both our model (Fig.~\ref{fig:fig4}(a)) and empirical data (Figs.~\ref{fig:fig4}(b)\,-\,(d)) clearly show two bright areas characterized 
by a high density of users with likely-minded neighbors, identifying two echo chambers
corresponding to opposite opinion groups.


In conclusion, we proposed a simple model that combines network and opinion dynamics,
and reproduced crucial features of empirical social networks characterized by polarization 
and echo chambers. 
The model is based on three  main assumptions
inspired by empirical evidence: i) aggregated social influence, 
ii) heterogenous activity, and iii) homophily in the interactions. 
While the role of social influence in opinion polarization has been 
extensively studied, the effect of opinion reinforcement and controversialness remains poorly understood, 
and has only recently started to be 
addressed \cite{Garimella:2018:QCS:3178568.3140565}. Within our model it is 
identified as one of the main features driving the transition between global 
consensus and radicalization.
In the case of controversial issues, a reinforcement 
mechanism leads to radicalization dynamics  
and may drive groups of agents away from the global consensus.
For weak homophily the transition from consensus to 
radicalization dynamics can be predicted analytically.
It is important to remark that our model is based on a minimal number of assumptions. Thus it does not take into account some features of empirical social networks which might additionally drive polarization phenomena, such as targeted advertising or different credibility of users.
We hope that our work stimulates empirical research
on the dynamics of polarization in online social networks to support 
our claims about the interplay of 
homophily, controversialness and the reinforcement of opinions.



\begin{acknowledgments}
This work was developed within the
scope of the IRTG 1740/TRP 2015/50122-0 and funded
by the DFG and FAPESP. The authors are indebted to K. Garimella, G. De Francisci 
Morales, A. Gionis, and M. Mathioudakis for sharing Twitter 
data, and to Philipp H\"ovel, G. De Francisci Morales, L. Cerekwicki 
and F. Sagues for helpful comments and discussions.
\end{acknowledgments}

\bibliographystyle{apsrev4-1}

%

\pagebreak
\widetext
\begin{center}
\textbf{\large Supplemental Material}
\end{center}

\section{Numerical simulations}
In the following we give a detailed algorithmic description of the numerical implementation of our model.\\
At the beginning of each simulation we set the parameters of the model including  time step $dt$,
total number of agents $N$, homophily $\beta$ and the reciprocity $r$ and define the initial conditions $x_i(t_0)\in[-1,1]$. Moreover we fix the parameters of the AD dynamics ($m=10$, $\epsilon=0.01$, $\gamma=2.1$) and draw activity values $a_i$ individually for each agent from the power law distribution $F(a) = \frac{1-\gamma}{1-\epsilon^{1-\gamma}}\,a^{-\gamma}$.\\

In each time step of the numerical algorithm the opinions and the temporal matrix are updated in the following way:
\begin{itemize}
    \item[1)] Each agent $i$ is activated with probability $a_i$\,. 
    \item[2)] If active: agent $i$ influences $m$ distinct agents $\{j\}$ chosen according to Eq.~(2)\,. \\This influence is expressed by a directed link $(j\rightarrow i)$ in the temporal adjacency matrix,\\ i.e. $A_{ji}(t_n)=1$, $\forall j$.
    \item[3)] With probability $r$ the directed link $(j\rightarrow i)$ is reciprocal, such that there is also a directed \\ link $(i\rightarrow j)$. 
    Hence, agent $i$ receives influence (social feedback) from $j$, i.e. $A_{ij}(t_n)=1$\,.
    \item[4)] Opinions $x_i$ are updated by numerically integrating Eq.~(1) using $A_{ij}(t_n)$\,.
    \item[5)] After each time step $t_n$ the temporal network $A_{ij}(t_n)$ is deleted.  
    
    \end{itemize}

We integrate the system of Eqs.~(1) using an explicit fourth-order Runge-Kutta method \cite{press2007numerical} and a time step of $dt=0.01$. This leads to a timescale separation between the AD dynamics and the opinion evolution of a factor of 100.

\newpage
\section{Polarized opinion distributions}
The opinion distributions $P(x)$ of all three investigated datasets
(\textbf{obamacare}, \textbf{guncontrol} and \textbf{abortion})
show two pronounced maxima on both sides of the neutral consensus. 
For sufficiently high values of $K$ and/or $\alpha$ 
the bimodular shape of the empirical distributions 
is reproduced by the model. 
This is, however, only the case if additionally homophily 
is introduced $(\beta>0)$.

\begin{figure}[h!]
\centering
\includegraphics[width=\linewidth]{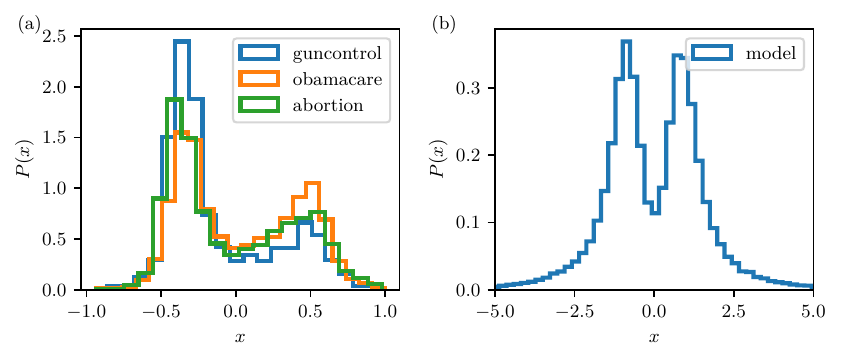}
\caption{Normalized opinion distributions as obtained from three different empirical data sets (a) and by simulating the model (b). 
For sufficiently high values 
of the parameters $K$, $\alpha$ and $\beta$ (here $K=3$, $\alpha=3$,
$\beta=1$, $r=0.65$) the model enters a polarized state characterized by bimodal opinion distributions, which are in qualitative agreement with the investigated Twitter data.}
\label{SMfig:polarized-opinion-states}
\end{figure}

\newpage
\section{Lifetime of polarized states}
\begin{figure}[h!]
\centering
\includegraphics[width=0.9\linewidth]{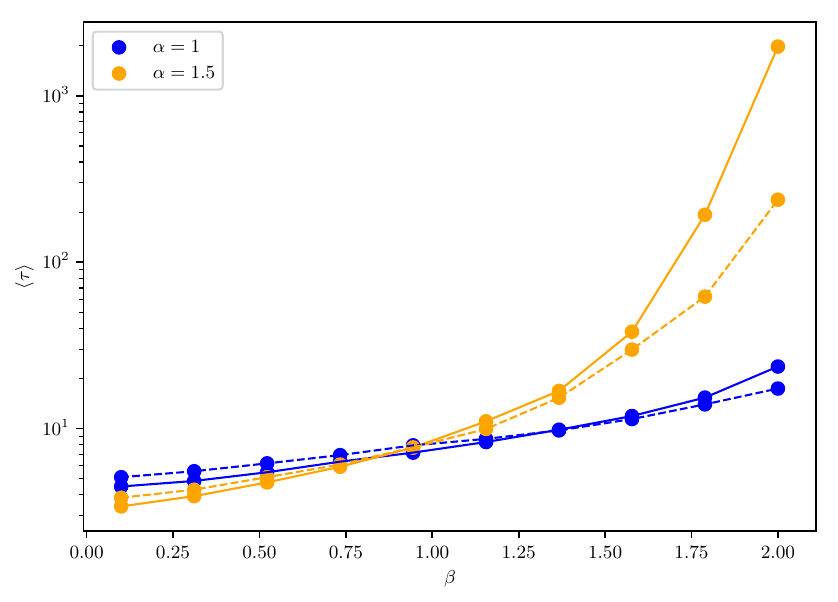}
\caption{The mean life-time of the polarized state strongly increases with the value of $\beta$. Each dot 
corresponds to the average of 1000 simulations 
of populations of $N=250$ agents with $dt=0.05$. The colors
correspond to different values of the controversialness $\alpha$, while $K=1$ for all depicted curves. The solid and dotted lines correspond to cases of $r=1$ and $r=0.5$, respectively. }
\label{SMfig:lifetimes}
\end{figure}

Polarized opinion states
will eventually decay into one-sided radicalized states, cf. Fig.~1(b) and Fig.~1(c) of the main text, respectively. However, their lifetimes $\tau$ 
strongly increase with the value of $\beta$. 
In Fig.~\ref{SMfig:lifetimes} we depict the mean lifetime, $\langle \tau\rangle$, 
as a function of $\beta$ for two different values of the 
controversialness $\alpha$ in the case of perfectly 
symmetric interactions ($r=1$) and temporal networks with reduced reciprocity ($r=0.65$).
Note the logscale on the 
$y$-axis, i.e. the strong dependence of the mean lifetimes on $\beta$, which even exceed an exponential growth for higher values of homophily.

The qualitative explanation of this effect is as follows. After the symmetric initialization of opinions $x_i$, each agent $i$ finds relatively fast its quasiequilibrium (metastable) opinion. In the absence of homophily these local opinions rapidly relax to a global one on either side of the opinion spectrum. The introduction 
of homophily, however, drastically changes the picture. 
While the agents with similar opinions interact frequently and form a 
metastable phase, 
agents with strongly diverging  
opinions hardly communicate. This results in two phases on opposite sides of the neutral 
consensus ($x_i=0$), which are practically
non-interacting and therefore long-living.

\section{Approximation of the critical controversialness $\alpha_c$}
Fast network dynamics allow the adjacency matrix $A_{ij}(t)$ (cf. Eq.~(1) of the 
main text) to be approximated 
by its time average, yielding
\begin{equation}\label{eq:mean_field_model}
 \dot{x}_i = -x_i + K \sum_{j=1}^N \langle A_{ij}(t)\rangle_t \tanh(\alpha x_j)\,.
\end{equation}
Neglecting homophily $(\beta=0)$ the overall probability that agent $i$ is influenced by $j$ has two different contributions. Either, $j$ contacts $i$, or  $i$ contacts $j$ and the link is reciprocal. Those two different processes happen with probabilities of $\frac{m}{N}a_j$ and $\frac{m}{N}a_i r$, respectively.
For the time averaged $ij$-th element of the adjacency matrix we therefore get in total
\begin{equation}\label{eq:mean-field-A_ij}
    \langle A_{ij}(t)\rangle_t = \frac{m}{N}(ra_i+a_j)\,,
\end{equation}
which, averaged over all activities in the system, becomes
\begin{equation}\label{eq:mean-field-A_ij2}
   \Lambda = \frac{m}{N}(1+r)\langle a\rangle\,.
\end{equation}
For a power law activity distribution $F(a) = \frac{1-\gamma}{1-\epsilon^{1-\gamma}}\,a^{-\gamma}$, normalized on the interval 
$a\in[\epsilon,1]$, we have
$\langle a\rangle=\frac{1-\gamma}{2-\gamma}
\frac{1-\epsilon^{2-\gamma}}{1-\epsilon^{1-\gamma}}$. 
Combining Eq.~\eqref{eq:mean_field_model} and Eq.~\eqref{eq:mean-field-A_ij2} 
 we get
\begin{equation}\label{eq:mean_field_model2}
 \dot{x}_i = -x_i + K\Lambda \sum_{j=1}^N \tanh(\alpha x_j)\,.
\end{equation}
To study the transition from neutral consensus 
to radicalization dynamics within this mean-field approach
we compute the Jacobian matrix of the system of Eqs.~\eqref{eq:mean_field_model2}, yielding 
\begin{equation}
\left.  \mathbb{J}\right\rvert_{\mathbf{x} = 0}=
\begin{bmatrix}
    -1 & K\Lambda\alpha &  \dots  & K\Lambda\alpha \\
    K\Lambda\alpha & -1 &  \dots  & K\Lambda\alpha \\
    \vdots & \vdots & \vdots & \vdots \\
    K\Lambda\alpha & K\Lambda\alpha & \dots  & -1
\end{bmatrix}\,,
\end{equation}
where all off-diagonal elements equal $K\Lambda\alpha$. 
The largest eigenvalue of $\mathbb{J}$ reads
\begin{equation}
\tilde{\lambda} = (N-1)K\alpha\Lambda-1=\frac{(N-1)}{N} K\alpha m(1+r)\langle a\rangle-1
\end{equation}
and determines the stability of the fixed point $\mathbf{x}=0$ 
with respect to small perturbations. For 
$\tilde{\lambda}>0$ the neutral consensus destabilizes, hence, 
$\tilde{\lambda}=0$ defines the critical value of controversialness $\alpha_c$,
i.e.
\begin{equation}
    \alpha_c= \frac{N}{(N-1)}\frac{1}{(1+r)Km\langle a\rangle}\,.
\end{equation}
In the limit of $N\rightarrow\infty$, Eq.~(3) of the main text is recovered.

\newpage
\section{Empirical data sets}
The datasets used in this work have been collected, analyzed and validated in previous works \cite{Garimella:2018:PDS:3178876.3186139,512b65372a05457fae9af34647985bb4, Garimella:2018:QCS:3178568.3140565}.
We use three different datasets from Twitter, each
of which contains a set of tweets on a given controversial topic of discussion: \textbf{abortion}, \textbf{obamacare}, \textbf{guncontrol}.
In order to keep the three datasets independent, we exclude users present in more than one dataset. 
In Ref. \cite{Garimella:2018:PDS:3178876.3186139}, the authors performed simple checks to
remove bots, using minimum and maximum thresholds for the number
of tweets per day, followers, friends, and ensure that the account
is at least one year old at the time of data collection.

Each dataset is built by collecting tweets posted during specific
events that led to an increased interest in the respective topic, during a time
period of one week around the event (3 days before and 3 days
after the event).
Users with less than 5 tweets about the issue during this time window were discarded.
The final numbers of users ($N_\mathrm{u}$) 
and measured reciprocities ($r$) for each data set are:\\ \textbf{abortion}: $N_\mathrm{u}=4130$, $r=0.69$,  \textbf{obamacare}: $N_\mathrm{u}=4828$, $r=0.62$, \textbf{guncontrol}: $N_\mathrm{u}=1838$, $r=0.61$. 

In Ref. \cite{Garimella:2018:PDS:3178876.3186139}, for each dataset, the directed follower network
among users has been reconstructed: a directed link from node $u$ to node $v$ indicates that user $u$ follows user $v$.
For each user, a political leaning score is inferred on the basis of the content s/he produces, by using a ground truth of political leaning scores of various news organizations (e.g., nytimes.com) obtained from Bakshy et al. [4]. 
Specifically, each news organization is classified by a score which takes values between 0 and 1.  A value close to 1 (0) indicates that the domain has a conservative (liberal) bent in their
coverage.
 From this classification, the political leaning score, or opinion, of each user $i$ is reconstructed by considering all tweets posted by user $i$ that contain a link to an online news organization with a known political leaning. Each tweet is thus associated with an opinion, corresponding to the political leaning of the news organization linked. 
 The political leaning of the user $i$ is defined as the average of the opinions expressed in his/her tweets. 
 Note that we transformed the original political leaning inferred in Ref. \cite{Garimella:2018:PDS:3178876.3186139}, from 0 to 1, into a score from -1 to 1, for coherence with the model.

\newpage

\section{Relation between user opinions and activities}
\begin{figure}[h!]
\centering
\includegraphics[width=\linewidth]{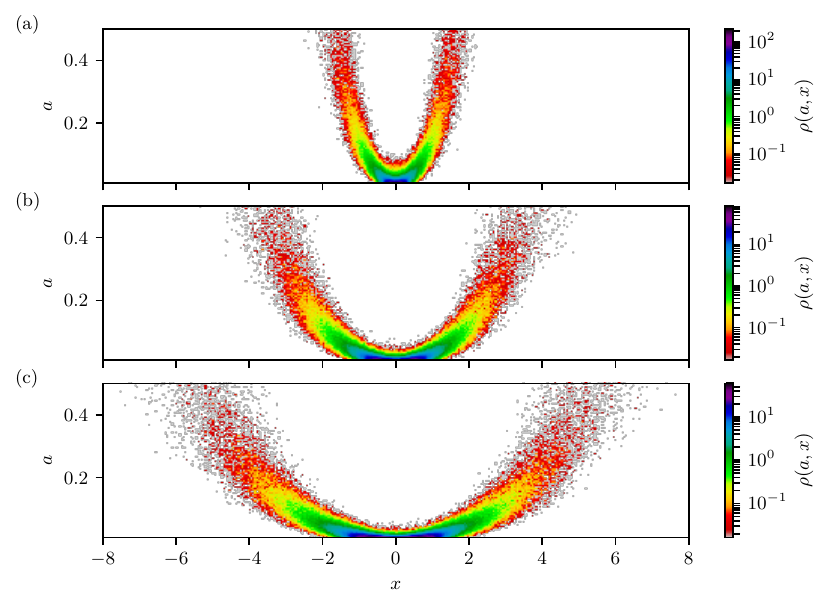}
\caption{Normalized histograms of simulation results 
in $x$-$a$ space which
depict the relation between opinions $x$ and activities $a$ 
of agents in a polarized state. The color encodes the density of agents.
U-shapes for increasing values of the social interaction strength are depicted from top to bottom ($K=1,2,3$), while we fixed the remaining parameters $\alpha=3$, $\beta=1$ and $r=.65$.}
\label{fig:SMfig2}
\end{figure}

The U-shaped relation between opinions $x$ and 
activities $a$ is a generic feature of the radicalization 
dynamics and occurs as soon as the system is in a polarized
state.
As an example in Fig.~\ref{fig:SMfig2} we vary the social 
interaction strength $K$ from top to bottom, while 
leaving all other model parameters constant as in Fig.~(3)b of the main text.
For increasing values of $K$ the convictions of 
agents of similar activities are increased. This results in a flattening of the U-shaped relation between activities and opinions.

\newpage
\section{Robustness with respect to reciprocity}
\begin{figure}[h!]
\centering
\includegraphics[width=1.\linewidth]{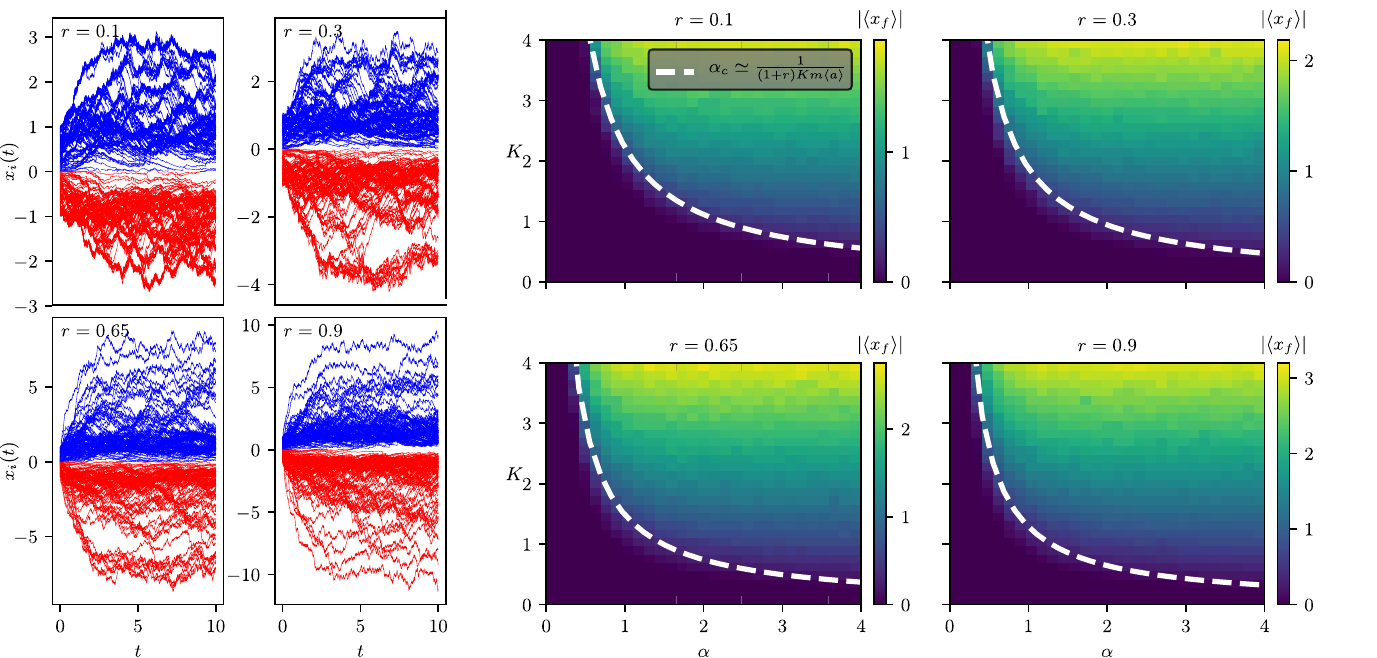}
\caption{Polarized states and transition to radicalization in $K-\alpha$ space for different values of the reciprocity $r\in[0.1,0.3,0.65,0.9]$. The remaining model parameters are identical to those of Fig.~(1)c and Fig.~(2), respectively.}
\label{SMfig:r-expl-fig12}
\end{figure}

As we mention in the main text the behavior of the model 
is remarkably robust with respect to the reciprocity of the
temporal networks $r$. To demonstrate this we depict the 
results of all main findings for additional values of $r$. In Fig.~\ref{SMfig:r-expl-fig12} we show the results of the model parameterized as in Fig.~(1)c and Fig.~(2), respectively, for additional four values of the reciprocity. 

First, and most 
importantly, we find that the emergence of polarized states does not rely on high values of $r$. Even in the case of $r=0.1$ the opinions clearly split in two groups on opposite sides of the 
neutral consensus. Note, that opinions reach lower absolute values for decreasing reciprocity. This is due to the fact that highly active agents get radicalized less, as they attract less social feedback from their
contacted peers. This effect is also reflected in the transition from consensus to radicalization dynamics, which happens already for lower 
values of $K$ and $\alpha$ for high reciprocities.
Intuitively, this makes sense: for low reciprocity (and low social feedback) agents do not self-radicalize so strongly and therefore tend to "forget" their opinion and strive towards zero, i.e. $x_i\rightarrow 0$. 
Note, that the mean-field approximation of this transition also works well in the case of low reciprocities.

\begin{figure}[h!]
\centering
\includegraphics[width=1.\linewidth]{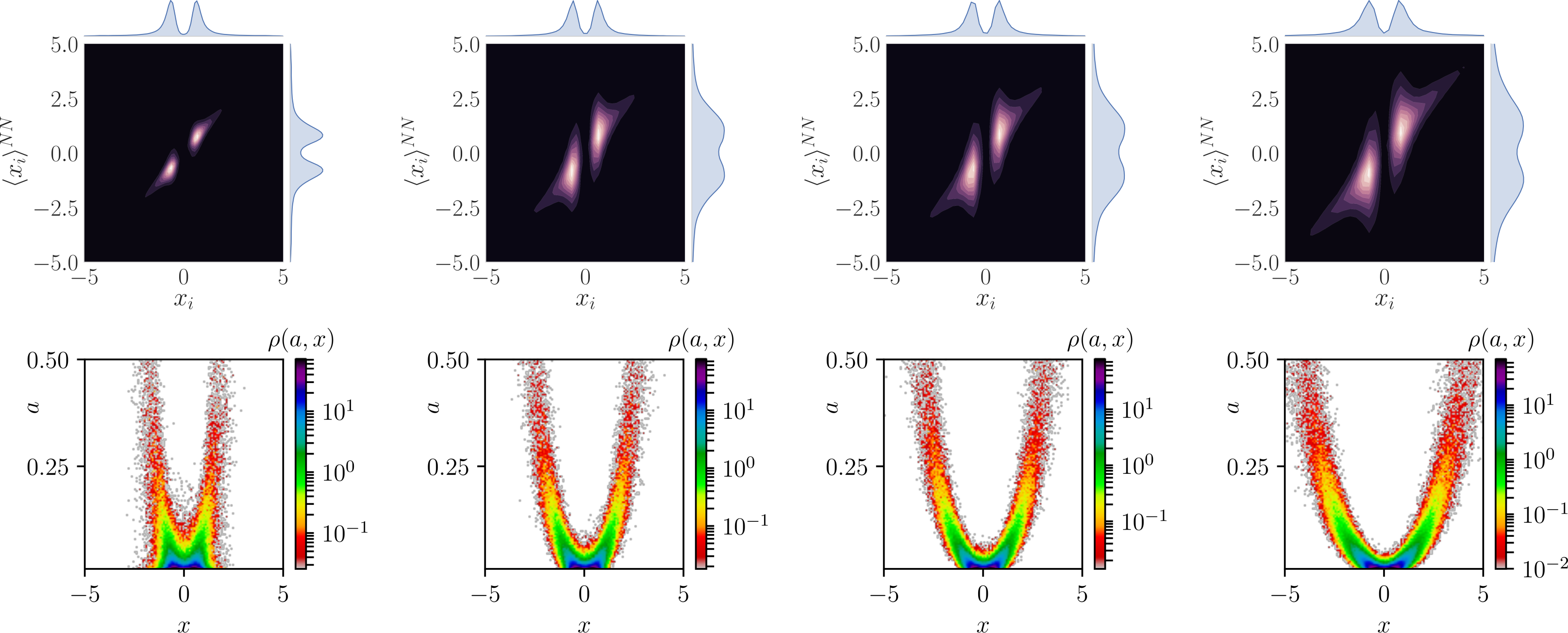}
\caption{Echo chambers and activity - opinion relation for different values of the reciprocity $r\in[0.1,0.3,0.5,0.9]$ from left to right. The remaining model parameters are identical to those of Fig.~(3) and Fig.~(4), respectively.}
\label{SMfig:r-expl-fig34}
\end{figure}

\newpage
To contrast the empirically measured Twitter data sets with the model, reciprocity was set to $r=0.65$, cf. Figs.~(3,4) of the main text. In Fig.~\ref{SMfig:r-expl-fig34} we show that the results obtained there are also robust for different values of $r$. Interestingly, echo chambers do not disappear 
even for very small values of the reciprocity.

\newpage
\section{Robustness with respect to asymmetric initial conditions and the effect of fluctuations}

In the main text we initialize the model uniformly spaced on the interval $x_i\in[-1,1]$ as a natural choice for initially balanced distributions of opinions. Here, we demonstrate that the model does not critically depend on this choice and the emergence of consensus and opinion polarization is recovered also for highly asymmetric cases. Furthermore, we briefly discuss the effects of fluctuations on radicalized states. To formalize the problem, we introduce a parameter $\delta$ shifting the initial interval of opinions, $x_i\in[-1+\delta, 1+\delta]$ to tune its asymmetry. The case of $\delta=0$ corresponds to the symmetric situation considered in the main text. 

\begin{figure}[h!]
\centering
\includegraphics[width=\linewidth]{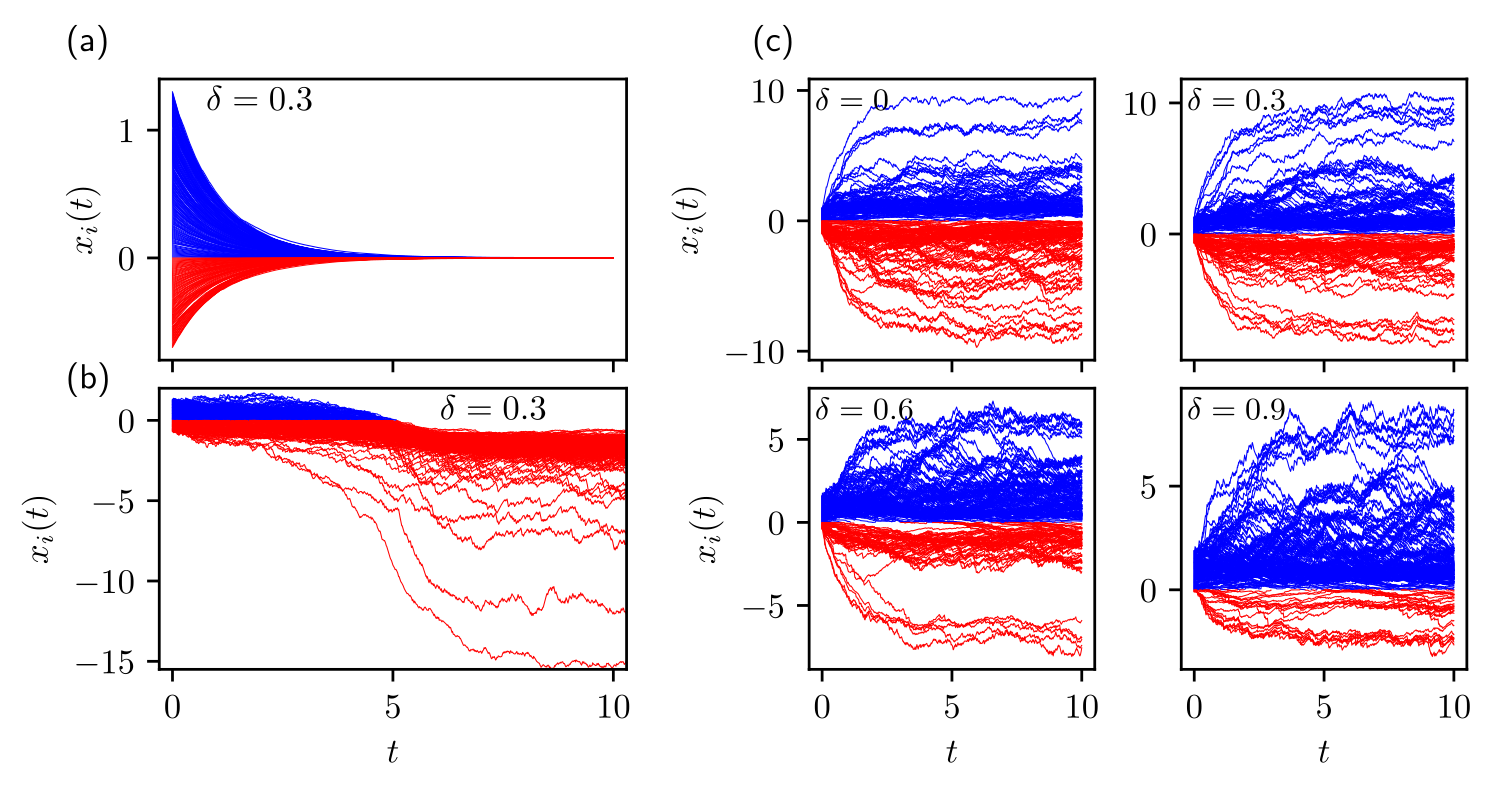}
\caption{Opinion evolution for symmetric and asymmetric initial conditions. The parameters of the model in (a)-(c) are set as in the corresponding subpanels of Fig.~1(a)-(c) of the main text. As in the case of symmetric initial conditions the model gives rise to consensus (a), (one-sided) radicalization (b) and opinion polarization (c). The degree of asymmetry is denoted by the parameter $\delta$.}
\label{SMfig:asymmetric_initial_conditions}
\end{figure}

In Fig.~\ref{SMfig:asymmetric_initial_conditions}(a) we show the emergence of consensus for $\delta=0.3$. Just as in the case of symmetric initial conditions (cf. Fig.~1(a) of the main text) the system reaches a full consensus. Note, that this behavior does not depend on the value of $\delta$ and results in the same final state even with all opinions initialized on one side, i.e. for $\delta<-1$ or $\delta>1$. 

For (one-sided) radicalized states this is not the case. Here, we observe that, due to higher values of $K$ and $\alpha$, the system is strongly governed by fluctuations. Those are essentially fluctuations of the network interaction strength due to the fast switching dynamics of the AD model and play a role similar to that of thermal noise in the dynamics of magnets. 
The influence of such fluctuations strongly depends on the initial conditions. While for symmetric initial conditions the system is purely fluctuation driven and both final states, $\sigma(x_i)\pm=1$, are realized with equal probabilities, initial conditions with $\delta\neq0$ favor final states with $\sigma(x_i)=\sigma(\delta)$. In such situations, cases like the one depicted in Fig.~\ref{SMfig:asymmetric_initial_conditions}(b) are observed less frequently.  Here, the initial conditions are biased towards the positive stance ($\delta=0.3$) and yet the system is absorbed by the state $\sigma(x_i)=-1$. 

In Fig.~\ref{SMfig:asymmetric_initial_conditions}(c) we show four runs of opinion polarization with increasing values of $\delta\in[0,0.3,0.6,0.9]$. Apart from the initial conditions, the model is parameterized as shown in Fig.~1(c) of the main text and shows the same qualitative behavior. Even in the case of strongly asymmetric initial conditions $(\delta\in[0.6,0.9])$ the radicalization mechanism leads to the emergence of a persistent polarized state.

\bibliographystyle{apsrev4-1}

\end{document}